# Understanding Student Pathways in Context-rich Problems

Pavlo D. Antonenko, John Jackman, Piyamart Kumsaikaew, Rahul R. Marathe,
Dale S. Niederhauser, C. A. Ogilvie[a], and Sarah M. Ryan
Iowa State University, Ames, IA 50011

In this paper we investigate the extent to which students' problem-solving behaviors change as a result of working on multi-faceted, context-rich problems. During the semester, groups of two to three students work on several problems that require more than one concept and hence cannot be readily solved with simple "plug-and-chug" strategies. The problems are presented to students in a data-rich, online problem-solving environment that tracks which information items are selected by students as they attempt to solve the problem. The students also complete a variety of tasks, such as entering their qualitative analysis into an online form. Students are not constrained to complete these tasks in any order. As they gain more experience in solving multifaceted physics problems, the student groups show some progression towards expert-like behavior: earlier qualitative analysis and more selective requests for information. However, there is room for more improvement as approximately half of the groups still complete the qualitative analysis task towards the end of the solution instead of earlier when it would be most useful to their work.

## I. Introduction

Progress in our technological society absolutely requires that young scientists and engineers have strong problem-solving skills that enable them to address new challenges that are often ill-defined and open-ended[1]. Students, who are used to working on well-structured, algorithmic problems, struggle when confronted with the multiple challenges of more complex tasks. Typically they either approach the problem by searching for an algorithm that might work ("plug-and-chug"), or freeze completely and ask immediately for direct help, or flounder by doing a large amount of busy-work with no real planning or direction.

We do not yet know which pedagogical methods best help our students develop the skills needed to solve ill-structured problems. However, insight can be obtained from research on the differences between experts and novices[2] which has documented how these two groups of people approach ill-structured tasks. Experts who successfully approach ill-structured problems have strong, organized conceptual knowledge[2,3] in the domain, which allows them to qualitatively analyze problems first[4,5] in order to rapidly get to the main essence of the problem and not be distracted by surface features, or fine details that will not be needed till later on in the solution. Experts also have strong, more general, metacognitive skills, such as monitoring the progress of their solution and worrying if the solution is still going in the correct direction[4], as well as evaluation skills such as knowing the importance of testing the solution against the assumptions during the solution process, using extreme conditions to check the solution's validity etc. Strong problem-solvers also know the benefit of incorporating the experience gained from each problem into their knowledge-structure that they can draw on when confronted with the next problem[6].

---

[a] Corresponding author, cogilvie@iastate.edu

It is feasible that successful pedagogies for ill-structured problem-solving will draw from this expert/novice research as well as building on instruction methods that help students solve more well-structured problems[7,8]. A common trait in these latter methods is to break the problem down into several steps and to explicitly teach these steps to students. Multistep procedures been developed in many domains such as, mathematics[9], chemistry[10], and biology[11]. In physics, perhaps the first reported use in the literature was from Reif[12], where the explicitly taught steps were summarized as Description, Planning, Implementation, and Checking. This was extended by Halloun and Hestenes[13], who showed that student solving performance improved with guided practice on these explicit problem-solving strategies.

This pedagogy is taken a step further when a tangible space is provided for students to follow the explicit multistep procedure. For example Active Learning Problem Sheets (ALPS) devised by van Heuvelen[14] contain separate, identified, sections where students must represent the problem graphically and develop a qualitative analysis before working on the mathematical representation. There are also sections on evaluation of units and magnitude of the answers. This approach has been transformed to a computer environment, Hierarchical Analysis Tool (HAT), in which students are constrained to first choose the principles involved in the problem from a pull-down menu, followed by associated concepts, then finally the equations they might be able to use[15, 16]. This tool has been shown to help students both categorize problems and improve their problem-solving performance.

A similar environment is provided by the Story Problem-Solving Environment (SPSE[17]). A story problem is presented to students who must then follow a series of tasks, ranging from identifying the principles involved, to qualitatively analyzing the problem, to building a quantitative representation of the problem.

In most of these examples students worked on well-structured problems. Hence two questions are whether the students use these techniques when faced with more ill-structured problems and whether they revert to their preferred plug-and-chug techniques once the scaffolds of the environments are removed or reduced. For example, flexibility in student access was recommended by Baker and O'Neil[18] in order to increase the fidelity and validity of online measures of complex problem solving.

We address both questions in this paper by observing how students solve moderately ill-structured problems in an online environment that does not constrain when students complete different tasks in solving the problem. The key goal of this paper is to establish the extent to which students move from novice to expert-like behavior as the result of guided practice in moderately ill-structured problems.

### A. Multifaceted context-rich problems

Multifaceted problems lie somewhere between well-structured problems and large, ill-defined, open-ended challenges. These problems require that students integrate *multiple* concepts in building a solution[19]. Typically these problems are context-rich, that is, they place the student in the middle of a challenge, for example, "you are a design engineer for a company that has been asked to build a ski ramp." However, the main characteristic is that the problems involve more than one concept, hence students cannot readily use a direct algorithmic approach ("plug and chug") as in a classical textbook exercise. As an example of a thermodynamic problem:

> *You are in charge of drinks at a picnic that will start at 3pm. You place ice inside a cooler at 6am, when the temperature outside is 10ºC. The day is forecast to warm up steadily to reach 30ºC by 3pm. Estimate how much ice you will need.*

At least two concepts are involved: heat transfer through a wall, and the amount of heat required to melt ice. It is also moderately ill-structured, in that the problem statement does not specify the wall thickness of the cooler, or the material used. Students must identify that they need these quantities for a final solution and then find that information. Although the problem is not mathematically complex, it does require that students as a group need to discuss the problem, identify the main concepts that are involved, qualitatively analyze the problem, find or estimate the required information and from there build a solution.

Multifaceted problems have been advocated by several groups across many disciplines for both school and university use, for example, the physics education research groups at University of Minnesota[19] and Ohio State University[20], in chemistry[21], industrial engineering[22,23,24], and across several disciplines by the Cognition and Technology Group at Vanderbilt[25], and the IMMEX project at UCLA[26,27,28].

### B. Tracking student pathways through a complex problem

The problem-solving environment we describe in this paper presents students with a general description of the task at hand, a menu of items that contain information, and a list of forms to fill out as they make decisions about how to solve the problem. This environment is similar to the IMMEX online tool[26] in listing information that students may or may not need to solve the problem, but it also contains written tasks (qualitative analysis, verification and cross-checks) that the students complete as they solve the problem. The environment does not constrain the students in the order in which they request information about the problem, nor the tasks that they complete. Students are free to make decisions about how they approach the problem, with the conjecture that if expert-like decisions produce student success, then these decisions may be more likely "owned" by the students and used in future contexts. The research advantage of the online tool is that we can track student pathways through these problems and establish the extent to which their problem-solving changes as a result of pedagogical interventions.

We build on the work done by Stevens[27] with IMMEX who showed that students start with a scattershot approach to requesting information in an unconstrained environment, but with experience they develop more expert-like strategies. Stevens and co-workers[27,28] used a neural-net to group the solution strategies of high-school students solving chemistry and biology problems into a few characteristic approaches: (1) the Prolific strategy, whereby a student requests a broad range of both relevant and irrelevant information, (2) the Redundant strategy, in which a student requests information that s/he already has, e.g., in the case of a diagnosis-case, the student orders a test that provides information on only an already eliminated diagnosis, (3) the Efficient strategy whereby a student requests the pertinent information, and (4) the Limited strategy, in which a student makes a guess at the solution without having requested sufficient information. As students solved more problems, they moved away from the Prolific strategy to the Efficient strategy. Similar improvements were noted by Chung and Baker[29] as their students solved several design tasks in an online environment.

We extend this work by also examining when students report tasks they completed while solving the problem. In particular, when do students describe their qualitative analysis of the problem: early in the solution when it may be most useful, or after they have found a solution when it may be easier to complete this task? We also cross-check the information-gathering result from Stevens and co-workers[27,28] by examining which data resources the students request and when. Do they follow a "prolific" strategy of collecting all available data early in the solution, or do they realize that for complex problems there is so much relevant and irrelevant data that this technique becomes less useful, and hence move to a strategy of first figuring out what they need and select only that information?

There are pros and cons to using online tracking of student pathways to explore how students solve problems. Manually observing and recording students in interviews provides much more information per student group, but it is expensive and limited to few groups. It is also difficult to have sufficient interviews to observe the progress of students throughout a semester. Online tracking increases the size of the data sample, however introduces complications such as the time when students complete their description of a task via the webform can be well after when they actually did that work. We address this concern later in the paper. A further advantage of using online tracking is that the data are collected in context, i.e. in a normal class environment and not in a lab setting.

## II. Data collection

**A. Educational Context**
The data presented in this study come from the Spring 2006 semester of a sophomore, calculus-based physics course at Iowa State University. Three hundred and fifty students took the course that was taught by one of the authors of this paper (CO). The course met for three lectures each week, one recitation and one lab. The active-learning format of the lecture was approximately 10 minutes of mini-lecture about an idea, followed by a conceptual question (referred to as a concepTest[30]) which the students answered via infrared-clickers; first individually, followed by a group discussion, and then recommitting as a group. The recitations used a mixture of Physics Tutorials[31], and context-rich, multi-faceted problem solving[19] designed to increase problem-solving skills.

Each topic in the course followed approximately the same sequence of an introductory tutorial during recitation to address the main concepts, two to three lectures, a lab, and two problem-sets. The first problem-set was due early in this sequence and contained mainly conceptual questions. The second problem-set focused on standard end-of-chapter problems designed to reinforce the basic procedural knowledge in the topic area.

The context-rich problems served as a capstone event for the topic area. Groups of two to three students (termed a "team" in this paper) worked on these problems during their recitation session where approximately 20 students meet with their teaching-assistant (TA). The TA's role is to provide guided instruction on how to qualitatively analyze the problems, i.e., to work from concepts and diagrams to build a solution. The TAs were trained to use leading prompts in all their discussions with students. At the start of the semester, this is a challenge for the TAs because their inclination is to provide more direct help to the students, for example, to make a suggestion of an approach or to identify the key constraint in a problem. We train the TAs how to scaffold: to ask prompts[32] that support the students in the early stages of the semester; such as
> *What information is missing? How are … related*
> *to each other? What do you think are the primary*
> *factors of this problem? Why is it …? Please explain.*

Emphasis was also placed on helping students to monitor their solution as it progressed. Schoenfeld[4] was one of the first to identify the difference between novices and experts in ongoing monitoring of their work. He analyzed many tape-recordings of students and faculty solving mathematics problems. Experts would periodically stop to check if the solution was making good progress, whether it was consistent with the original plan, or whether the algebra was getting too messy. Novices (and our students), on the other hand, would plow on regardless of how the solution was progressing.

During the semester the students worked on five context-rich problems; two in thermodynamics, one in waves, and two in magnetism and magnetic induction. The problem descriptions are given in Appendix A and more details on the problems can be provided upon request to the authors. Due to restrictions in space approximately half of the recitation sections solved the problems in a pen-and-paper format, and half used the online Problem-solving Learning Portal (PSLP) environment[22,23,24] to access the problem with its information and to submit their solution. The PSLP environment is described below. In this paper we use the tracking information from PSLP to analyze how students approach these problems, in what order do the students gather information, analyze the problem, and identify the principles involved before submitting a solution.

These recitation sections met in a class-room that had one computer per team of two-to-three students. The makeup of the teams was not actively managed by the TA; some teams stayed together during the semester, others reformed and students did not attend every recitation. The login page for PSLP required one student to start the problem, and others could then add their name. However, because not all teams signed-in their complete group, for this paper we will treat the teams as anonymous and look at the overall behavior of the class.

**B. Problem-solving environment**

The PSLP was developed by the authors to present ill-structured story problems in an on-line environment. PSLP's shell can be populated with problem-scenarios in different content areas. An initial screen presents the problem statement to the students (Figure 1), a menu bar across the top includes the tasks students complete as they solve the problem, and a menu bar on the left lists resources, tools, and pre-packaged advice. Students determine the sequence in which they fill-in the tasks and access resources.

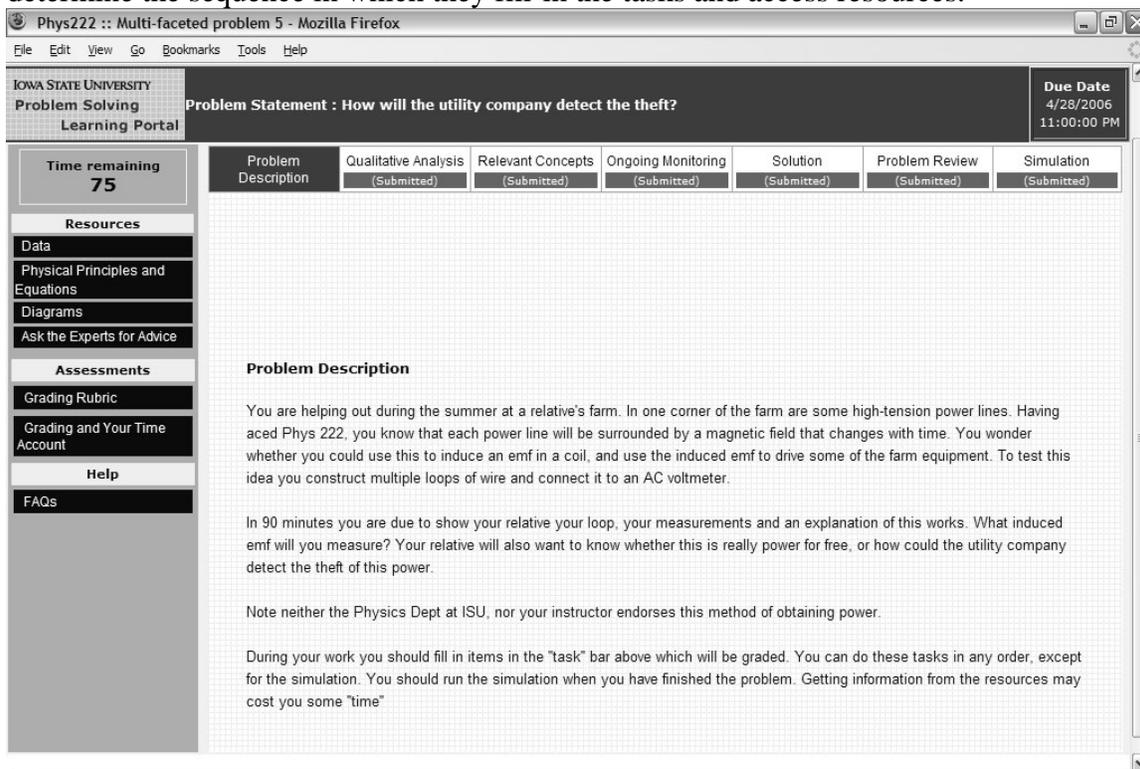

**Figure 1: Screen-shot of entry-page into PSLP with problem-description**

Resources were split into four categories: data, physical principles, diagrams, and advice. Each category contained several pieces of information, some of which were relevant and some irrelevant to the problem. Accessing each resource opened a new window — allowing participants to view several windows on their screen — but accessing resources also cost students "time" from their "time-remaining" account (top-left of screen). The problems were written in the second-person, i.e. "You are helping out at a relative's farm for the summer…" and were posed as a challenge to be met before meeting with another person in 90 "minutes" time. Accessing resources reduced this "time" account. Each piece of information typically cost between 5 and 10 "minutes".

Near the top of the screen was a task-bar where students select a task and enter their work. These tasks included
- Qualitative analysis: textual information written by students that describes qualitatively what processes take place in the problem.
- Relevant concepts: a set of check boxes where students select concepts that directly apply to the solution of this problem.
- Ongoing monitoring: a text box where students describe what checks they have made during their solution.
- Solution: a set of radio buttons with typically 15 different numerical values as possible answers to the problem.
- Problem review: a text box where students describe what checks they have made after they obtained their solution.
- Simulation: a Macromedia Flash simulation of the problem that was only available to students after they had submitted their solution. The students were warned that once the simulation was run they would not be able to change any of their answers.

Students could complete these task descriptions at any time in the problem and complete these tasks in any order. Resubmission of these tasks was allowed.

## III. RESULTS
### A. Resource Gathering
One essential aspect of successful problem solving is collecting and extracting relevant information from the resources at hand. This section discusses how students access information as they progress through five physics problems (i.e., from the beginning to the end of the semester). Table 1 presents basic information about the number of teams, total resources, and number of relevant resources for these five problems.

| Problem | Number of Teams | Number of Resources Available | Number of Relevant Resources |
|---|---|---|---|
| 1 | 54 | 20 | 11 |
| 2 | 42 | 18 | 9 |
| 3[b] | 22 | 19 | 8 |
| 4 | 30 | 21 | 9 |
| 5 | 36 | 24 | 9 |

**TABLE 1. Number of teams, total resources available, and relevant resources in multifaceted problems 1-5**

---

[b] Technical problems in the first few recitations reduced the sample size for this problem.

Analysis of the system tracking data demonstrated that student teams tended to use only about 25% of the resources available to them in problems 1 through 5. Overall, the fraction of resource pages accessed, $f_{resource} = \frac{resources\ accessed}{available\ resources}$, while solving problems decreased from problem 1 ($f_{resource}$ = 0.29) to problem 5 ($f_{resource}$ = 0.22). Furthermore, in problem 3, one team of students didn't open any resources during their problem-solving process, and in problem 5 the number of such teams increased to seven. These teams did not tend to submit correct solutions, however. These teams may correspond to the "Limited" strategy class identified by Stevens et al.[27] : in which students make a guess at the solution without having requested sufficient information.

As the teams progressed through the semester, they displayed several distinct patterns of accessing information. For the first problem, many teams tended to open resource pages in the order that they were presented to them on the screen (i.e., top to bottom), viewing both relevant and irrelevant resources. Figure 2 shows a typical click-map for resource requests. Each row corresponds to a team and the boxes represent the sequence of requests for resources for problem 1. Relevant and irrelevant resources accessed by the students were color-coded.

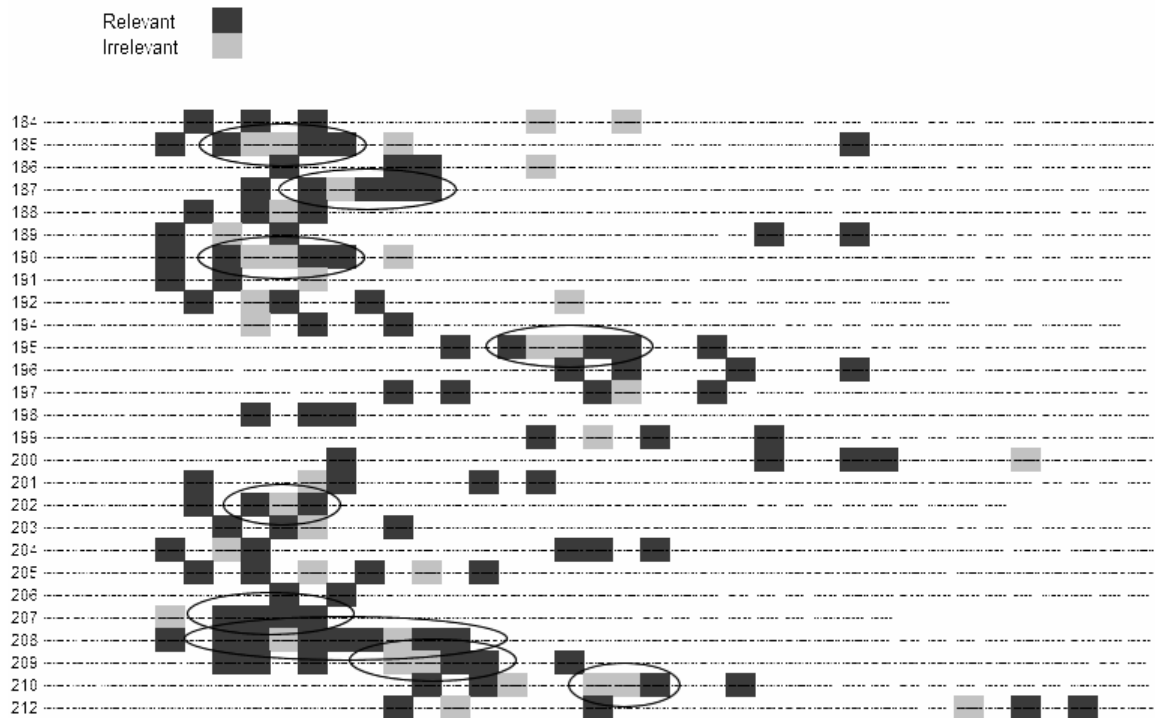

**FIG. 2. Click map of teams' problem-solving sessions for problem 1. Each row corresponds to a team and the boxes represent the sequence of requests for resources. Dark boxes correspond to requests for relevant resources while light boxes are requests for irrelevant resources. The circles show the "clustered" pattern of accessing resources.**

A "cluster" of requests is defined as an uninterrupted sequence of 3 or more resource requests. If a team has a cluster of requests, then they are not interrupting the resource gathering process and going to the task pages. This pattern changed, however, as the semester progressed. The percentage of teams who exhibited a "clustered" pattern of accessing resources (Figure 1) was 39% in problem 1. This decreased significantly in problem 2 (24% of teams), and disappeared almost completely in problem 3 (5% of teams), problem 4 (3% of teams), and problem 5 (3% of teams). This result confirms what Stevens et al.[27] found with students working in the IMMEX environment: that students moved from "prolific" gatherers of many resources in a sequence, to a behavior that intersperses resource gathering with other tasks.

A positive correlation was found between whether a team accessed resources as clusters and $f_{resource}$ for a team. For problem 1 this correlation was very strong: $r = 0.552$ and $p = 0.01$. Thus, the likelihood that students would view resources in clusters increased with the amount of resources that were opened by the students.

The fraction of relevant resources that were accessed, $f_{relevant} = \frac{relevant\ resources\ accessed}{relevant\ resources\ available}$, starts to differentiate the behavior of teams. A negative correlation was found between $f_{relevant}$ and $f_{resource}$. For problem 1 this correlation was $r = -0.406$ with $p = 0.01$, i.e., more expert-like teams who accessed a larger fraction of the relevant resources, accessed a smaller fraction of the total resources.

A key question is whether more teams demonstrated this behavior as the semester progressed. Figure 3 demonstrates that the fraction of relevant resources opened averaged across the student teams increased from $\overline{f_{relevant}} = 0.78$ (problem 1) to 0.85 (problem 3), although the error on these means indicates that this difference is not significant.

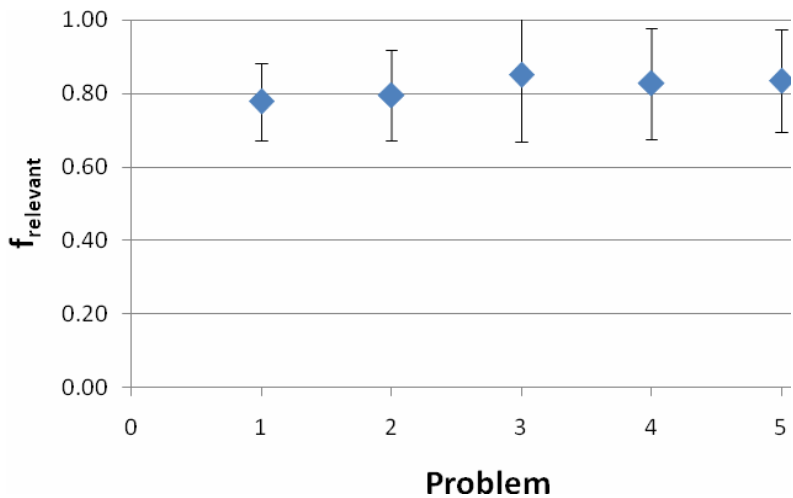

**FIG. 3. Mean fraction of relevant resources $\overline{f}_{relevant}$ for multifaceted problems 1-5. The error bars indicate the standard error on the mean.**

Another pattern of accessing information resources can be inferred from which type of information was viewed. Most of the resources available to students in PSLP were presented in two categories: a) data, and b) principles. We can characterize each team's behavior by $f_{data} = \dfrac{data\ resources\ accessed}{total\ resources\ accessed}$ and $f_{principles} = \dfrac{principles\ resources\ accessed}{total\ resources\ accessed}$. While the data resources comprised the bulk of information accessed by the teams across all five problems, the fraction of physics principles resources increased significantly in problems 3, 4, and 5 (Figure 4).

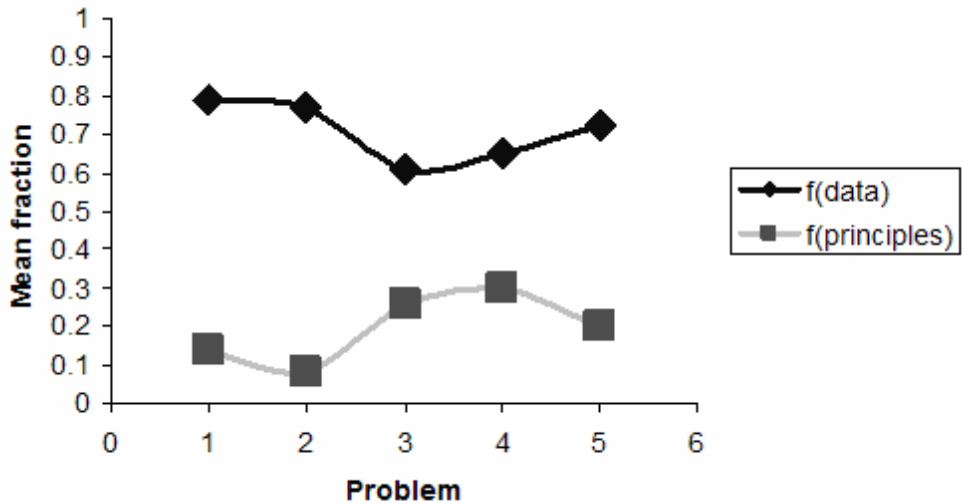

**FIG. 4. Mean fraction of data resources ($f_{data}$) and principles resources ($f_{principles}$) that student teams requested during multifaceted problems 1-5.**

Taken together, these results indicate that as students gain more experience in solving multifaceted physics problems increases, they tend to access more information on physical laws and principles, they are more likely to request relevant information, and these requests are less likely to come in uninterrupted clusters.

### B. Completion of Tasks

PSLP tracking data showed that as students were solving the problem, they tended to complete the tasks in the order that was presented to them via the task bar shown in Figure 1. Table 2 shows that the order of mean completion times was the same for all five problems. The table shows the average time of completing the tasks with time=0 corresponding to the beginning of each team's problem session.

| Problem | Qualitative Analysis (min) | Relevant Concepts (min) | Ongoing Monitoring (min) | Solution (min) | Problem Review (min) |
| --- | --- | --- | --- | --- | --- |
| 1 | 41 | 42 | 50 | 52 | 53 |
| 2 | 33 | 36 | 49 | 51 | 52 |
| 3 | 44 | 47 | 54 | 57 | 57 |
| 4 | 39 | 41 | 48 | 55 | 55 |
| 5 | 29 | 31 | 44 | 48 | - |

**TABLE 2: Mean time of completion of each task for multifaceted problems 1-5**

As noted in an earlier section, we record the time that the teams enter their analyses via web forms, which occurs after teams discuss the task. From the closeness of the times for ongoing monitoring and solution, it seems that the students are describing what they did for ongoing monitoring well after they may have done this task. In contrast, the submission times for the qualitative analysis and relevant concepts are separated from the solution time by 10 to 15 minutes, so students have still not finished working on the problem by the time they enter their description of these tasks.

It is notable that the average completion time of the qualitative analysis task decreased from 41 minutes in problem 1, to 29 minutes in problem 5. This persists even when normalizing by the time each group submitted their problem solution: $(t_{QualAnal}/t_{solution}) = 0.78$ for problem 1 and $(t_{QualAnal}/t_{solution}) = 0.61$ for problem 5, i.e., the qualitative analysis is completed 20% earlier in problem 5 than in problem 1.

Expert-like behavior is to conduct an early qualitative analysis of the problem and use this to determine which resources (data, principles) are needed to solve the problem. A more novice-like behavior is to first collect many of the resources (both relevant and irrelevant), work on the problem, and after the problem is largely solved, the team summarizes what the problem is qualitatively about. The idea that novices would complete a qualitative analysis as a summary task was suggested to the authors by K. VanLehn [33] due to the likelihood that there is a reduced cognitive load on novices to perform a qualitative analysis after they have spent significant amount of time on the problem. Savelsbergh[34] used this idea in a pedagogical approach where students re-represent the problem as a schematic sketch after the problem has been solved instead of before.

To explore this further we examined the relative time between when students completed the qualitative analysis section and when they requested resources. We computed the time when each team had reached 80% of their total resource requests and compared that with the time they completed the qualitative analysis. Both times were normalized for each team using the total problem solving time. Figure 5 presents a scatterplot for problem 1 of normalized time viewing 80% of resources versus the normalized time submitting qualitative analysis. Each point corresponds to a single team.

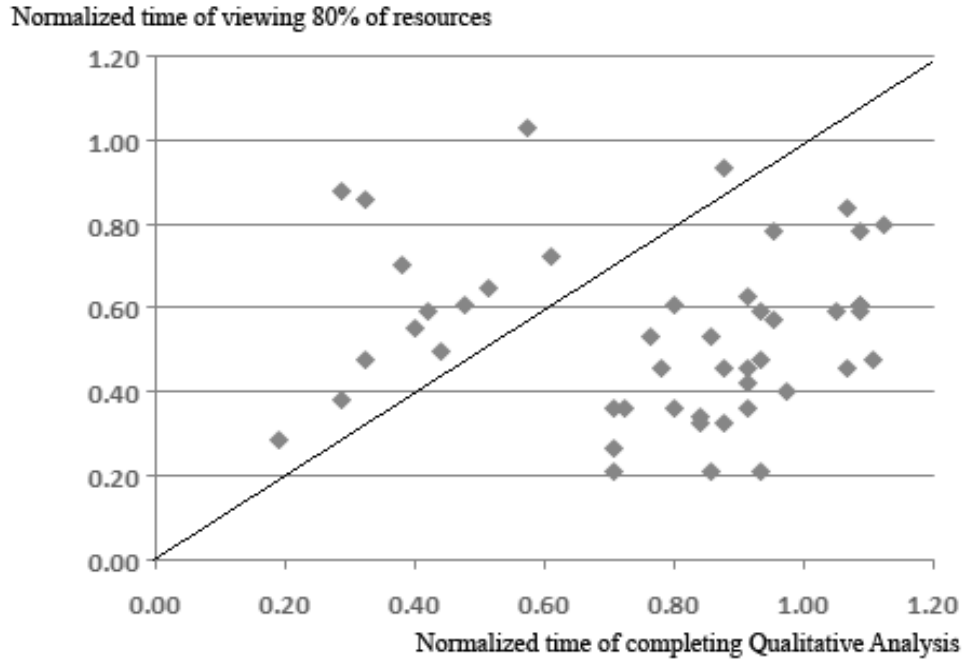

**FIG. 5. Normalized time each team had viewed 80% of the total resource requests versus the normalized time when each team submitted a qualitative analysis for problem 1. The times are normalized by the time that a team submitted their solution.**

In the top left region of this scatter-plot we find teams that completed their qualitative analysis before requesting 80% of their resources. This would be considered more expert-like behavior of using a qualitative analysis to inform which resources to gather. The bottom-right of this scatter-plot comprises teams that submitted their qualitative analysis after gathering 80% of their resources, and in some cases, after submitting their solution (normalized times > 1.). The lower-right group either found it cognitively easier to do their qualitative analysis after they had completed a large amount of work or they did this task earlier but only reported it just before entering their solution.

Clearly there are two classes of behavior separated by the diagonal line, with the more expert-like teams in the top, left section. A key question is the extent to which student teams change their behavior as the semester progresses. We calculated the percentage of teams who fell into this top, left section (expert-like) for each of the 5 problems, i.e., teams that completed qualitative analysis before they viewed 80% of the information resources. This fraction doubles from 26% in problem 1, to approximately half the teams for problems 3 to 4, and slightly decreases to 39% in problem 5 (see Table 3). The fraction of groups who entered a correct answer is also shown in this table: this fraction did not significantly differ between the two types of teams.

| Problem | Novice-like teams | Expert-like teams | Mean performance |
|---|---|---|---|
| 1 | 74% | 26% | 0.48 |
| 2 | 41% | 59% | 0.74 |
| 3[a] | 55% | 45% | 0.61 |
| 4 | 43% | 57% | 0.52 |
| 5[b] | 61% | 39% | 0.49 |

**TABLE 3: Problem-solving behavior and learning performance across the five multifaceted problems.**
[a] Skewed due to one missing value (one team not using any resources)
[b] Skewed due to seven missing values (seven teams not using any resources)

We also found that teams that submitted their qualitative analysis earlier tended to have a higher value for $f_{relevant}$ ($r = -0.389$, $p = 0.01$).

## IV Summary

In this paper we have described a problem-solving environment (PSLP) for multi-faceted problems. PSLP extends previous tools by having both a large collection of relevant and irrelevant resources available to students, as well as providing web forms for students to describe the tasks they are completing as they solve a problem. Students can complete these tasks or select resources in any order and hence are free to make decisions about how they approach a problem. We have tracked these choices via student pathways through these problems and have established the extent to which their problem-solving changes as a result of experience.

The problem-solving pedagogy used in this paper was introduced by Heller et al.[19] where students work on context-rich, multi-faceted problems with guided instruction from TAs. Groups of two to three students solve problems that involve more than one concept and hence can be more readily solved if students utilize more expert-like strategies of qualitative analysis and planning. These strategies are explicitly taught by the TAs, as well as ongoing monitoring and reviewing strategies.

As students gain more experience in solving multifaceted physics problems, they tend to access more information on physical laws and principles, they are more likely to request relevant information, and these requests are less likely to come in uninterrupted clusters. Student groups also complete their qualitative analysis of the problem 20% earlier in the last problem of the semester compared to the first problem, though this task is still completed by many groups after they have accessed most of the resources. The fraction of groups who complete the qualitative analysis task towards the end of the problem decreases from 75% for the first problem to close to 50% for the rest of the problems.

Taken together the student groups show some progression towards expert-like behavior: earlier qualitative analysis and more selective requests for information. However, there is room for more improvement: approximately half of the groups still complete the qualitative analysis task after they examine many resources.

This suggests a few possible options to improve the pedagogy of multi-faceted problems. The easiest to implement is for TAs to more actively encourage student groups to complete their qualitative analysis earlier. However, the TAs have responsibility for many groups during each session and they may not be able to provide groups with timely advice. One option is to automatically analyze the behavior of each student group in real-time and have PSLP provide hints, assistance to students [28], e.g. if the group is indiscriminately opening all the data resources, suggest that they perform a qualitative analysis of the problem. A further option is to provide feedback on the substance of the typed qualitative analysis with automatic text processing. Rose et al.[35] have started to use text-classification software that when trained on a large set of prior student answers can be used to give feedback in real-time to students when they enter their qualitative analysis.

The online problem-solving environment PSLP has great potential for providing this feedback to students as they develop their problem-solving skills. PSLP is also designed to be used in any discipline: the types of tasks and resources can be reconfigured for use in other subjects. Faculty interested in using the tool should contact the authors.

## Acknowledgements
This work was funded in part by grants from The HP Technology for Teaching grant program, the Iowa State University Computing Advisory Committee, and an Iowa State University Miller Award.

## Appendix A: Problem Descriptions
1. "How much ice will you need?" You are in charge of keeping the drinks cold for a picnic. You have a styrofoam box that is filled with cola, water and you plan to put some 0° ice in it. Your task is to buy enough ice to put in the box at 6am so that the temperature stays at 0°C until the picnic starts at 4pm. You don't want to buy too much ice because that means that you'll have less money to spend on food and other picnic items.
   How much ice will you need? You have 90 "minutes" to calculate the amount of ice, before your cousin picks you up to drive to buy the ice. Getting information from the resources (on left-hand panel) may cost you some 'time'. The resources will only cost you "time" when you first access them and the cost will be always indicated. Your score will depend partially on how much 'time' you have left in your account.
2. "Optimal Operating Conditions" You are an engineer designing a nuclear power plant. The core of the reactor is designed to operate at a temperature of $T_H$ and the cooling water is at a temperature of $T_C$. Your group has found that you may be able to reduce the cost of the plant considerably by using smaller engines to convert the heat from the reactor into work in the form of electrical energy. The design concept you have developed is to use a tank of liquid lithium as heat buffer to be held at temperature T between $T_H$ and $T_C$ and then use one generator to operate between the core and the lithium tank and another to operate between the tank and cooling water.

In 90 mins you are to present your idea to the engineering committee which is to decide whether a full scale engineering study of this design is to be undertaken. You need to develop the case for this design, including what temperature T for the lithium tank produces the greatest efficiency and how does this efficiency compare to the standard design.

3. "Making a perfect fifth" Your friend, an artist, has been thinking about an interesting way to display a new wind sculpture she has just created. In order to create an aural as well as visual effect, she would like to use the wires to hang the sculpture as sort of a string instrument. Her basic design involves vertically hanging two pieces of wire from two eye-hooks on the ceiling, then hanging the heavy sculpture from a horizontal bar from some point along the bar. The distance between two eye-hooks on the ceiling is the same as the total length of the horizontal bar.

   The aural effect that she would like to achieve is that when the wind blows across two vertical strings, they play a perfect fifth, i.e. the ratio of the frequencies of the two sounds is 3:2.

   Your friend tells you that she has been successful in hanging the sculpture but not in choosing the point along the bar to hang the sculpture giving the desired sound. Desperate for success, she knows you are taking physics and asks you for help.

   In 90 minutes you are due to meet her at the local coffee shop. What is your advice concerning the design of the sculpture. What notes will the two strings play with your design?

4. "Designing a blood-flow meter" You have a summer internship at a company that makes medical instruments. During medical surgeries, there is a need to measure the amount of blood flow through arteries that have been exposed by the surgery, but otherwise have not been cut. That is blood is still flowing through these arteries

   You know from your studies of biochemistry that blood contains a reasonable amount of both positive and negative ions. If you place a small magnetic field across the artery, then these moving ions would experience a magnetic force. Your company also manufactures a range of devices that can measure the electrostatic potential between two points.

   In 90 minutes you are due to meet with your boss. You need to sketch out a device that could provide the blood flow based on the measurement of the electrostatic potential across two points on the artery. Based on the model you develop, what electrostatic potential would you expect to observe? For the device to be practicable it needs to respond relatively quickly, so you should also estimate the order of magnitude of time it takes for the electrostatic potential to develop across two points on the artery.

5. "How will the utility company detect the theft" You are helping out during the summer at a relative's farm. In one corner of the farm are some high-tension power lines. Having aced Phys 222, you know that each power line will be surrounded by a magnetic field that changes with time. You wonder whether you could use this to induce an emf in a coil, and use the induced emf to drive some of the farm equipment. To test this idea you construct multiple loops of wire and connect it to an AC voltmeter.

   In 90 minutes you are due to show your relative your loop, your measurements and an explanation of this works. What induced emf will you measure? Your relative will also want to know whether this is really power for free, or how could the utility company detect the theft of this power.

   Note neither the Physics Dept at ISU, nor your instructor endorses this method of obtaining power.

---

[1] P.M. King and K.S. Kitchener, "Developing reflective judgment: Understanding and promoting intellectual growth and critical thinking in adolescents and adults", San Franscisco, Jossey-Bass (1994).

[2] M.T.H. Chi, P.J. Feltovich, and R. Glaser, "Categorization and representation of physics problems by experts and novices", Cognitive Science 5, 121-152 (1981).

[3] J.D. Novak and D.B. Gowin, "Learning how to learn", Cambridge: Cambridge University Press (1984).

[4] A. Schoenfeld "Mathematical Problem Solving," London: Academic Press, (1985)

[5] N. Reid and M.-J. Yang, "Open-ended problem-solving in school chemistry: a preliminary investigation", International Journal of Science Education, 24, 1313-1332 (2002).

[6] C. Bereiter and M. Scardamalia, "Surpassing Ourselves: an inquiry into the nature and implications of expertise", Chicago, Open Court Publishing (1993).

[7] D. Gabel (Ed.), "Handbook of research on science teaching and learning", New York: Macmillan, (1993).

[8] R. Taconis, B. Hout-Wolters "Systematic comparison of solved problems as a cooperative learning task", Research in Science Education 29, 313, (1999).

[9] G. Polya, "How to Solve It", 2nd ed, New Jersey, Princeton University Press, (1957).

[10] D.M. Bunce, & H. Heikkinen, "The effects of an explicit problem-solving approach on mathematical chemistry achievement", Journal of Research in Science Teaching, 23, 11 (1986).

[11] R.W. Hurst and M.M. Milkent, "Facilitating successful prediction problem solving in biology through application of skill theory", Journal of Research in Science Teaching 33, 541-552, (1996).

[12] F. Reif, J.H. Larkin, and G.C. Brackett "Teaching General Learning and Problem-Solving Skills" Am. J. Phys. 44, 212-217 (1976)

[13] I.A. Halloun and D. Hestenes, "Modeling instruction in mechanics. American Journal of Physics, 55, 455-462 (1986)

[14] A. Van Heuvelen "Overview, Case Study Physics", American Journal of Physics, 59, 898 (1991)

[15] R.J. Dufresne, W.J. Gerace, P.T. Hardiman, J.P. Mestre, "Constraining Novices to Perform Expertlike Problem Analyses: Effects on Schema Acquisition" Journal of the Learning Sciences 2, 307, (1992)

[16] W.J. Leonard, R.J. Dufresne, and J.P. Mestre, "Using qualitative problem-solving strategies to highlight the role of conceptual knowledge in solving problems" American Journal of Physics 64, 1495 (1996)

[17] D.H. Jonassen, "Learning to Solve Problems: An Instructional Design Guide, Pfeiffer Publishing (2004).

[18] E.L. Baker and H.F. O'Neil, "Measuring problem solving in computer environments: current and future states", Computers in Human Behavior 18, 609-622 (2002).

[19] P. Heller, R. Keith and S. Anderson, "Teaching problem solving through cooperative grouping. Part 1: Group versus individual problem solving", American Journal of Physics 60, 627 (1992).

[20] A. Van Heuvelen "Overview, Case Study Physics", American Journal of Physics, 59, 898 (1991)

[21] N. Reid and M.-J. Yang, "Open-ended problem-solving in school chemistry: a preliminary investigation", International Journal of Science Education, 24, 1313 (2002).

[22] S. Olafsson, M. Huba, J. Jackman, F. Peters, and S. Ryan. Information Technology Based Active Learning: "A Pilot Study for Engineering Economy", in Proceedings of the 2003 ASEE Annual Conference, June 22-25, Nashville, TN.

[23] S. Olafsson, K. Saunders, J. Jackman, F. Peters, S. Ryan, V. Dark, and M. Huba. "Implementation and Assessment of Industrial Engineering Curriculum Reform", in Proceedings of the 2004 ASEE Annual Conference, June 20-23, Salt Lake City, UT.

[24] S. Ryan, J. Jackman, F. Peters, S. Olafsson, M. Huba. The Engineering Learning Portal for Problem Solving: Experience in a Large Engineering Economy Class. The Engineering Economist 49, 1-20 (2004)

[25] Cognition and Technology Group at Vanderbilt, "The Jasper experiment: An exploration of issues in learning and instructional design", Educational Technology Research and Development, 40, 65 (1992).

[26] R Stevens, J. Ikeda, A. Casillas, et al., "Artificial neural network-based performance assessments", Computers in Human Behavior, **15**, 295-313 (1999).

[27] R. Stevens and J. Palacio-Cayetano, "Design and Performance Frameworks for Constructing Problem-Solving Simulations", Cell Biology Education 2, 162 (2003).

[28] R. Stevens, A. Soller, M. Cooper, and M. Sprang, "Modeling the Development of Problem Solving Skills in Chemistry with a Web-Based Tutor", Lecture Notes in Computer Science **3220**, 580 (2004). Springer-Verlag, Heidelberg Germany.

[29] Gregory K.W.K. Chung and Eva L. Baker, "An Exploratory Study to Examine the Feasibility of Measuring Problem-Solving Processes Using a Click-Through Interface", J Technology, Learning and Assessment **2**, (2) (2002).

[30] E. Mazur, "Peer Instruction: A User's Manual", Series in Educational Innovation. Upper Saddle River, NJ, Prentice Hall, (1997).

[31] L. McDermott and P.S. Schaffer, "Tutorials in Introductory Physics", Upper Saddle River NJ: Prentice Hall (2002).

[32] G.E. Xun, S.M. Land, "A conceptual Framework for Scaffolding Ill-structured Problem-solving Processes using Question Prompts and Peer Interactions", Educational Technolgoy Research and Development, 52, 5, (2004)

[33] K. VanLehn, private communication (2005).

[34] E. Savelsbergh, T de Jong M.G.M. Ferguson-Hessler "Situational Knowledge in Physics: The Case of Electrodynamics" J. Res. in Sci. Teaching 39, 928 (2002)

[35] C.P. Rose`, private communication (2006).